\documentclass[twocolumn]{aastex631}
\usepackage{booktabs}
\usepackage[version=4]{mhchem}
\usepackage{txfonts}

\usepackage{amsmath}
\usepackage{sidecap}

\begin{document}

\title{Evidence for ubiquitous carbon grain destruction in hot protostellar envelopes}
\email{nazari@strw.leidenuniv.nl}

\author[0000-0002-4448-3871]{Pooneh Nazari}
\affiliation{Leiden Observatory, Leiden University, P.O. Box 9513, 2300 RA Leiden, the Netherlands}

\author[0000-0002-1103-3225]{Beno\^{i}t Tabone}
\affiliation{Universit\'{e} Paris-Saclay, CNRS, Institut d'Astrophysique Spatiale, 91405 Orsay, France}

\author[0000-0002-2555-9869]{Merel L. R. van ’t Hoff}
\affiliation{Department of Astronomy, University of Michigan, 1085 S. University Ave., Ann Arbor, MI 48109, USA}

\author[0000-0001-9133-8047]{Jes K. J{\o}rgensen}
\affiliation{Centre for Star and Planet Formation, Niels Bohr Institute \& Natural History Museum of Denmark, University of Copenhagen, {\O}ster Voldgade 5–7, 1350 Copenhagen K., Denmark}

\author[0000-0001-7591-1907]{Ewine F. van Dishoeck}
\affiliation{Leiden Observatory, Leiden University, P.O. Box 9513, 2300 RA Leiden, the Netherlands}
\affiliation{Max Planck Institut f\"{u}r Extraterrestrische Physik (MPE), Giessenbachstrasse 1, 85748 Garching, Germany}

\submitjournal{ApJ Letters}

%% Note that the \and command from previous versions of AASTeX is now
%% depreciated in this version as it is no longer necessary. AASTeX 
%% automatically takes care of all commas and "and"s between authors names.

%% AASTeX 6.31 has the new \collaboration and \nocollaboration commands to
%% provide the collaboration status of a group of authors. These commands 
%% can be used either before or after the list of corresponding authors. The
%% argument for \collaboration is the collaboration identifier. Authors are
%% encouraged to surround collaboration identifiers with ()s. The 
%% \nocollaboration command takes no argument and exists to indicate that
%% the nearby authors are not part of surrounding collaborations.

%% Mark off the abstract in the ``abstract'' environment. 
\begin{abstract}

Earth is deficient in carbon and nitrogen by up to ${\sim}4$ orders of magnitude compared with the Sun. Destruction of (carbon- and nitrogen-rich) refractory organics in the high-temperature planet forming regions could explain this deficiency. Assuming a refractory cometary composition for these grains, their destruction enhances nitrogen-containing oxygen-poor molecules in the hot gas ($\gtrsim 300$\,K) after the initial formation and sublimation of these molecules from oxygen-rich ices in the warm gas (${\sim}150$\,K). Using observations of $37$ high-mass protostars with ALMA, we find that oxygen-containing molecules (CH$_3$OH and HNCO) systematically show no enhancement in their hot component. In contrast, nitrogen-containing, oxygen-poor molecules (CH$_3$CN and C$_2$H$_3$CN) systematically show an enhancement of a factor ${\sim} 5$ in their hot component, pointing to additional production of these molecules in the hot gas. Assuming only thermal excitation conditions, we interpret these results as a signature of destruction of refractory organics, consistent with the cometary composition. This destruction implies a higher C/O and N/O in the hot gas than the warm gas, while, the exact values of these ratios depend on the fraction of grains that are effectively destroyed. This fraction can be found by future chemical models that constrain C/O and N/O from the abundances of minor carbon, nitrogen and oxygen carriers presented here. 

\end{abstract}

%% Keywords should appear after the \end{abstract} command. 
%% The AAS Journals now uses Unified Astronomy Thesaurus concepts:
%% https://astrothesaurus.org
%% You will be asked to selected these concepts during the submission process
%% but this old "keyword" functionality is maintained in case authors want
%% to include these concepts in their preprints.

%% From the front matter, we move on to the body of the paper.
%% Sections are demarcated by \section and \subsection, respectively.
%% Observe the use of the LaTeX \label
%% command after the \subsection to give a symbolic KEY to the
%% subsection for cross-referencing in a \ref command.
%% You can use LaTeX's \ref and \label commands to keep track of
%% cross-references to sections, equations, tables, and figures.
%% That way, if you change the order of any elements, LaTeX will
%% automatically renumber them.
%%
%% We recommend that authors also use the natbib \citealt
%% and \citet commands to identify citations.  The citations are
%% tied to the reference list via symbolic KEYs. The KEY corresponds
%% to the KEY in the \bibitem in the reference list below. 

\section{Introduction} 
\label{sec:intro}

Molecules containing carbon are important in discussions related to the origin of life on Earth or other terrestrial planets (\citealt{Oberg2021}). However, the Earth is known to be deficient in carbon and nitrogen by at least ${\sim}1-2$ and up to 4 orders of magnitude compared with the interstellar grains and many comets including 1P/Halley and 67P/Churyumov–Gerasimenko (\citealt{Geiss1987}; \citealt{Allegre2001}; \citealt{Wooden2008}; \citealt{Marty2012}; \citealt{Bergin2015}; \citealt{Rubin2019}; \citealt{Fischer2020}). To explain this difference, multiple works have suggested that Earth should have formed from solids that are dominated by silicates and depleted of carbon (e.g., \citealt{Jura2012}; \citealt{Gail2017}; \citealt{Li2021}). 

Assuming that all grains start with composition similar to those of the interstellar grains (\citealt{Rubin2019}; \citealt{Duval2022}), a second step of processing of these grains should occur in the inner planet-forming regions around the young star to explain the depletion in carbon. This further processing is thought to be the destruction of solid material that are rich in carbon. Destruction of carbon-rich solids in the inner most regions is also implicated by the deficiency of carbon in the atmospheres of polluted white dwarfs (\citealt{Jura2006}, \citealt{Jura2007}; \citealt{Gansicke2012}; \citealt{Farihi2013}; \citealt{Farihi2016}) due to accretion of carbon-poor asteroids by the white dwarfs. 

\begin{figure}[ht!]
\plotone{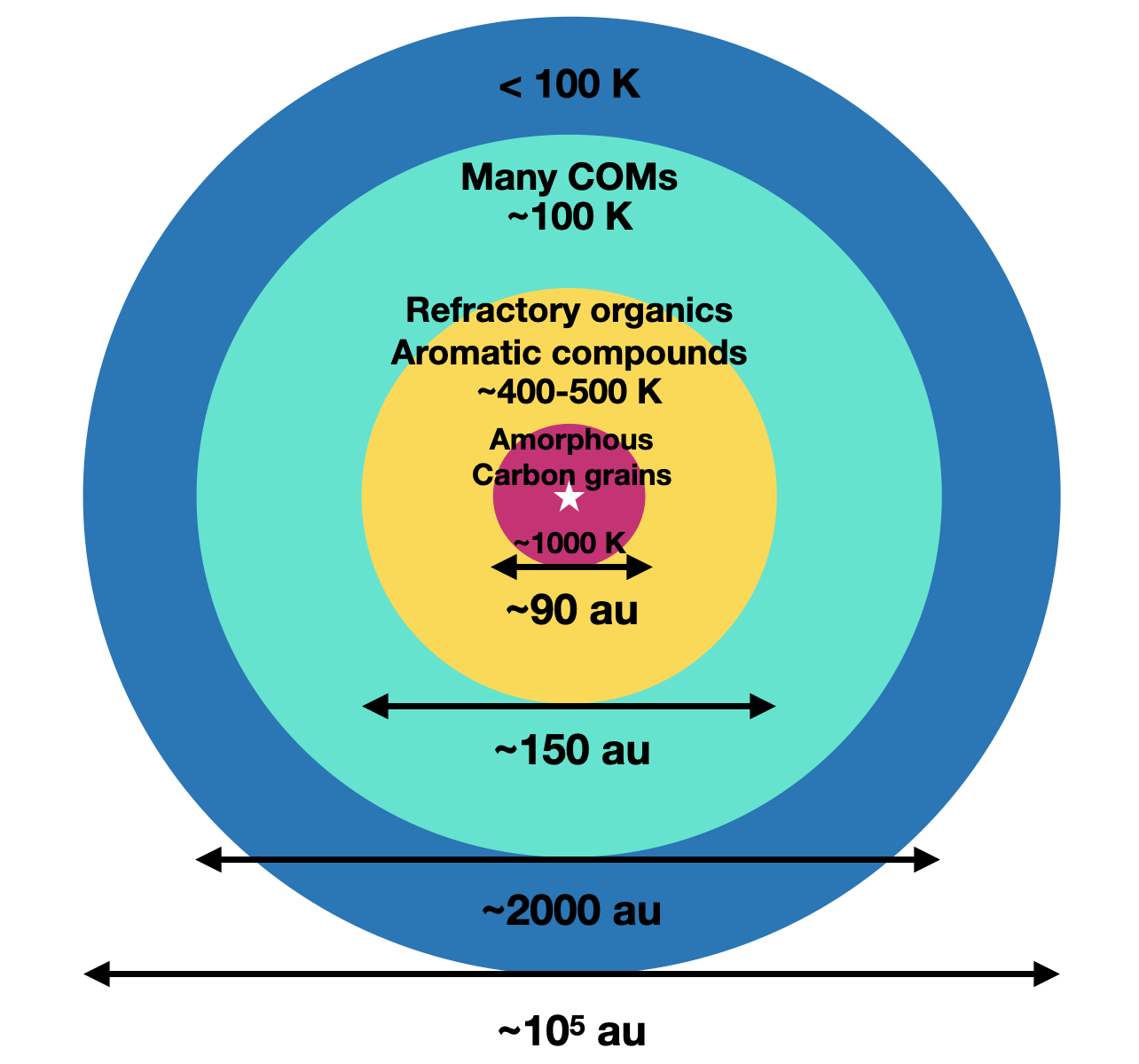}
\caption{The approximate temperatures at which various components from dust grains enter the gas in the envelope around a young star. At temperatures below ${\sim} 100$\,K (cold) all complex organic molecules (COMs) are in the ices. At temperatures above ${\sim} 100$\,K (warm) many COMs including CH$_3$CN and CH$_3$OH sublimate into the gas. At temperatures of around 400\,K (hot) aromatic compounds evaporate (\citealt{Gail2017}) and at similar temperatures refractory organics are thought to enter the gas through sublimation or pyrolysis (\citealt{Nakano2003}; \citealt{Gail2017}; \citealt{Li2021}). Finally, at temperatures of ${\sim} 1000$\,K (very hot) pure amorphous carbon grains are thought to get destroyed through oxidation (\citealt{Finocchi1997}; \citealt{Lee2010}; \citealt{Gail2017}) or UV photolysis (\citealt{Alata2014}; \citealt{Anderson2017}; \citealt{Klarmann2018}). The physical scale of each region is shown for a typical high-mass protostellar system. 
\label{fig:layers}}
\end{figure}

Most of the carbon is in the so-called refractory organic matter that in addition to carbon contains other elements such as H, N and O, while there are smaller fractions of carbon that exist in other forms such as hydrocarbons and pure carbon grains (\citealt{Kissel1986Giotto}; \citealt{Fomenkova1994}; \citealt{Fomenkova1997}; \citealt{Fomenkova1999}; \citealt{Bardyn2017}). These carbonaceous materials, depending on their form, are thought to be destroyed at different temperatures and through different mechanisms. Figure \ref{fig:layers} summarizes these temperatures as a simple cartoon (see the figure caption for references) ignoring any non-thermal sublimation processes. However, destruction of carbon grains has not yet been directly observed due to absence of spectral features and extinction in protostellar systems. Therefore, an indirect method to observe this phenomenon is to search for its effects on the chemical composition of the hot gas ($\gtrsim 300$\,K) located close to the young protostar.

Given that most of carbon is thought to be in refractory organics (\citealt{Kissel1986}; \citealt{Fomenkova1994}; \citealt{Fray2016}), here we consider the effect of destruction of these types of carbonaceous material on the hot gas chemistry. Destruction of refractory organics has been argued to happen within the first million years of the planetary system formation (\citealt{Li2021}). Furthermore, this destruction occurs at high temperatures (${\sim}400-500$\,K; \citealt{Li2021}; \citealt{Gail2017}). These high temperatures are achieved up to larger radii in the protostellar phase making young and embedded protostars the prime objects to search for any effect from carbon grain destruction on the gas-phase molecules. Moreover, recent ALMA results have shown that planets start forming in these earlier stages (\citealt{Manara2018}; \citealt{Tychoniec2020}) making the protostellar phase even more relevant for this study. 

In the protostellar envelope the temperature at which the refractory organics are sublimated could decrease from ${\sim}400-500$\,K to ${\sim}300$\,K due to the lower densities and pressures compared to those of the inner disk (see \citealt{vantHoff2020} for detailed explanation). Therefore, for spatially unresolved hot core observations where the disk and envelope are not separated the temperature at which the refractory organics are destroyed is likely between ${\sim}300$\,K and ${\sim}400$\,K. These hot regions are the focus of this work.

It has been speculated that destruction of the refractory organics will result in a top-down formation of oxygen-poor complex organic molecules (COMs, \citealt{Herbst2009}; \citealt{Ceccarelli2022}) in the hot gas (\citealt{vantHoff2020}). Therefore, both column densities and excitation temperatures of the oxygen-poor molecules should be affected if carbon grain destruction is effective. We note that this top-down gas-phase formation in the hot gas is an addition to the early cold phase bottom-up formation of COMs in ices (\citealt{Gibb2004}; \citealt{Jorgensen2020}) and the thermal sublimation of these species at temperatures of around 100\,K (see Figure \ref{fig:layers}). The hot gas-phase chemistry should be a natural result of destruction of refractory organics if the composition of refractory organics is similar to the refractory material of comets (\citealt{Rubin2019}). In particular, \cite{vantHoff2020} searched for differences in the excitation temperatures between O- and N-bearing COMs (i.e., molecules with and without oxygen) as a potential sign of sublimation of carbon grains. In their literature survey, they found that the excitation temperatures for the N-COMs are higher than those for the O-COMs in some sources and the opposite is true for some other sources. Given the non-homogeneous literature reports, while this study showed some interesting trends it was inconclusive, emphasizing the need for systematic surveys and analysis. 

In this work we use one such molecular surveys with the Atacama Large Millimeter/submillimeter Array (ALMA) to search for the signatures of destruction of carbon grains through a two-component temperature fit to oxygen-rich and oxygen-poor organics. A similar fitting method was performed by \cite{Neill2014} for their \textit{Herschel} observations. We consider two of the most abundant COMs, namely methanol (CH$_3$OH) and methyl cyanide (CH$_3$CN), in addition to two other molecules with and without oxygen in a large sample of high-luminosity (and potentially high-mass) protostellar systems. These additional species are isocyanic acid (HNCO) and vinyl cyanide (C$_2$H$_3$CN). For the purposes of this paper, when referred to oxygen-rich molecules we imply HNCO and CH$_3$OH and when referred to oxygen-poor molecules we imply CH$_3$CN and C$_2$H$_3$CN. We use data from the large program ALMA Evolutionary study of High Mass Protocluster Formation in the Galaxy (ALMAGAL, PIs: S. Molinari, P. Schilke, C. Battersby, P. Ho) because of its high sensitivity to detect both optically thin minor isotopologues of the most abundant molecules such as methanol to probe the warm gas and the optically thin weak lines of the major isotopologues with $E_{\rm up} > 400-500$\,K to probe the hot gas. Therefore, it is only now possible to measure the hot ($\gtrsim 300$\,K) and warm ($ 200 \gtrsim T_{\rm ex} \gtrsim 100$\,K) components of the above molecules from their high- and low-$E_{\rm up}$ (upper energy level) lines for a large sample of sources. 

In this letter we use data from the ALMAGAL survey to measure the hot column densities of HNCO, CH$_3$OH, CH$_3$CN, and C$_2$H$_3$CN. These molecules have similar sublimation temperatures. We use the detected lines of these molecules with $E_{\rm up} > 400-500$\,K (i.e., probing $T \gtrsim 300$\,K) for the hot components. We take the warm column densities for the above species from \cite{Nazari2022ALMAGAL}, who measure them from the $^{13}$C or $^{18}$O isotopologues of the targeted molecules which only have detected lines with $E_{\rm up} \lesssim 300-400$\,K (i.e., probing $T \sim 150$\,K). We then compare the ratios of the hot number of molecules to warm number of molecules for species with and without oxygen in a sample of 37 protostellar systems. We develop a simple analytical model to interpret the observations. Through the comparison of observations with the model we search for signs of destruction of refractory organics and find it to be common.

\section{Observations and methods}
\label{sec:obs_methods}
\subsection{Data}
\label{sec:data}
In this work we use the publicly available ALMA data from the ALMAGAL large program (project ID: 2019.1.00195.L). We consider the (high-luminosity and potentially high-mass) sources that were analyzed by \citet{Nazari2022ALMAGAL} (see \citealt{vanGelder2022} and \citealt{vanGelder2022Deuteration} for results on methanol). The works mentioned above present the full details of the data, here we only give a summary of the observational parameters.

All data used here have an angular resolution between ${\sim} 0.5\arcsec$ and ${\sim} 1.5\arcsec$. The angular resolution for high-mass sources located at distances of $2-3$ kpc correspond to radii of between ${\sim}500$\,au and ${\sim}2500$\,au. The frequency ranges covered for the selected sample are $217.00-218.87$\,GHz and ${\sim}219.07-220.95$\,GHz. The spectral resolution is around $0.7$\,km\,s$^{-1}$ which results in spectrally resolved molecular lines toward the protostellar systems. The spectra are the same as those used in \cite{Nazari2022ALMAGAL} which were extracted from the peak pixel of the CH$_3$CN 12$_4$–11$_4$ moment zero map. The line rms is ${\sim}0.2$\,K for the spectra.

\subsection{Spectral fitting}
\label{sec:spec_fit}
In this work we consider CH$_3$CN, CH$_3$OH, HNCO, and C$_2$H$_3$CN in the data described in Section \ref{sec:data} and analyze them with a two-component temperature model. The warm column densities of CH$_3$CN, CH$_3$OH and HNCO are measured from their $^{13}$C or $^{18}$O isotopologues in \cite{Nazari2022ALMAGAL}. The isotope ratios in that work were calculated from the relations given in \cite{Wilson1994} and \cite{Milam2005} and the distances of the objects from the Galactic center. If an object is an outlier with respect to those relations, the calculated column densities from isotope ratios could have an additional uncertainty. Using the isotopologues to calculate the column densities is desirable because it avoids the optically thick lines of the main isotopologues. However, it is limited to lower K lines and derivation of a single excitation temperature between 100\,K and 200\,K. In other words, the $^{13}$C or $^{18}$O isotopologues of CH$_3$CN, CH$_3$OH and HNCO only have lines with $E_{\rm up} < 400$\,K detected and hence those column densities miss the hot gas. 

Here we attempt to find the column densities of the relevant molecules in the hot regions closer to the protostars by only fitting the optically thin lines of the main isotopologues with $E_{\rm up} \gtrsim 400$\,K. The measured hot column densities and excitation temperatures for all the 37 sources are given in Table \ref{tab:results}. The line lists are taken from the Cologne Database for Molecular Spectroscopy (CDMS; \citealt{Muller2001}; \citealt{Muller2005}). Frequencies, quantum numbers, $E_{\rm up}$ and $A_{\rm ij}$ of the considered lines are given in Tables B.1 and E.7 of \cite{vanGelder2022} and \cite{Nazari2022ALMAGAL}. For more information on the spectroscopy see Appendix \ref{sec:spec_info}. We use the CASSIS\footnote{\url{http://cassis.irap.omp.eu/}} spectral analysis tool (\citealt{Vastel2015}) to fit the spectra assuming the same emitting area that was used in \cite{Nazari2022ALMAGAL} for the warm component. However, we stress that the relevant quantity for our study is the total number of molecules ($\mathcal{N} = A N$, where $A$ and $N$ are the emitting area and column density) which does not change if a different emitting area is used as long as the lines are optically thin. In other words, if a smaller emitting area is assumed the derived column density will increase and vice versa such that the total number of molecules stays the same. Therefore, after the derivation of column densities we compare the total numbers of hot and warm molecules with each other (see Section \ref{sec:obs_comparison}). For all sources the column density of the hot component is found using the same fit-by-eye method used in \cite{Nazari2022ALMAGAL} (also see \citealt{Chen2023}). Below we explain our method in more detail for each molecule.

\begin{figure*}[ht!]
\plotone{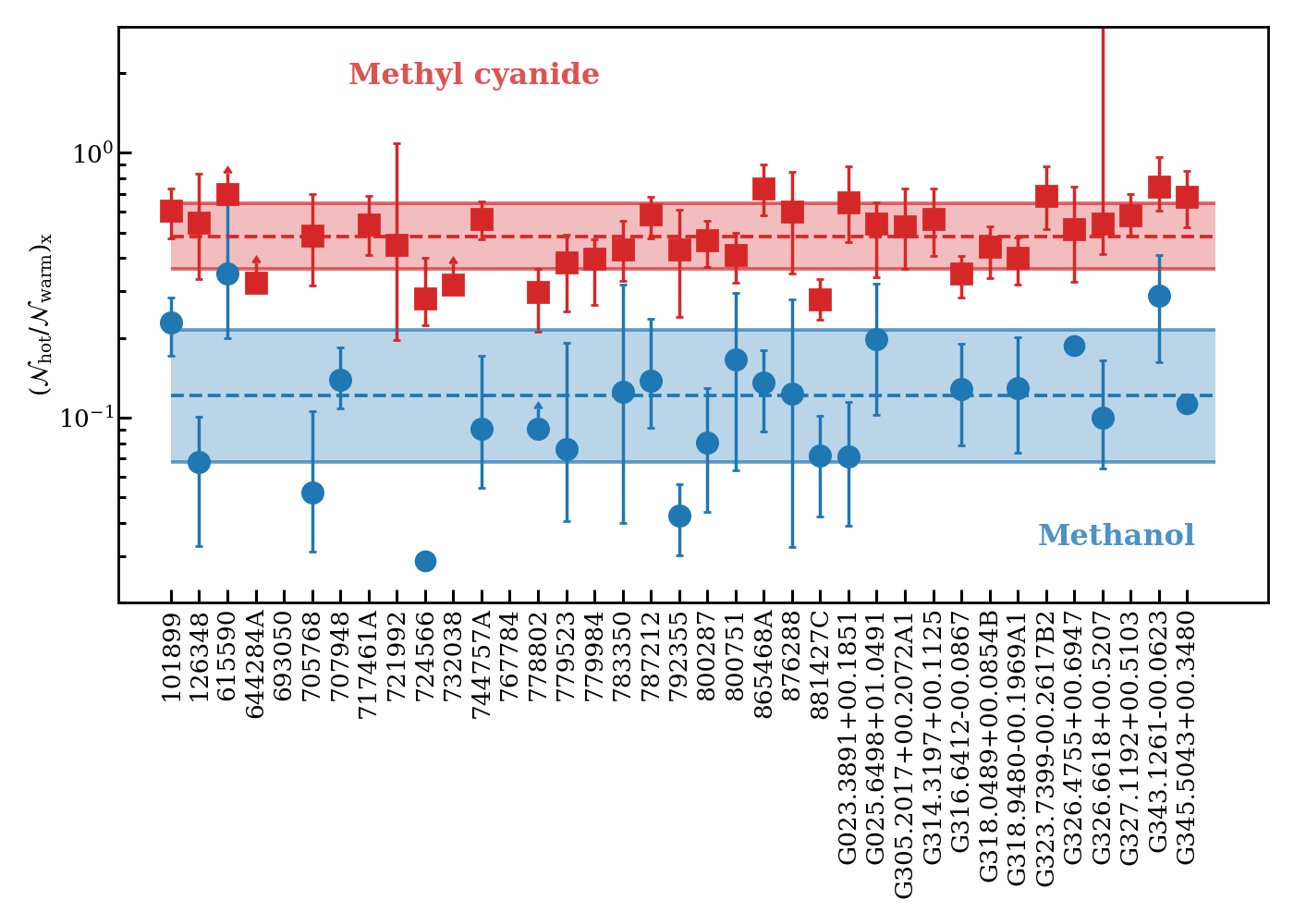}
\caption{Ratios of hot to warm number of methanol (blue circles) and methyl cyanide (red squares) molecules for the sources considered in this work. The dashed red and blue lines indicate the weighted mean (by errors) of $log_{10}$ of methyl cyanide and methanol data points. The shaded regions show the respective weighted standard deviation (by errors) of $log_{10}$ of the ratios around the mean. The values for the warm components are taken from \cite{Nazari2022ALMAGAL} for these sources and are calculated from the $^{13}$C and $^{18}$O column densities of methyl cyanide and methanol, respectively. There is an enhancement of a factor ${\sim} 4-5$ in the hot component of methyl cyanide compared to methanol.
\label{fig:ratio_methanol_methyl}}
\end{figure*}

Methyl cyanide (CH$_3$CN) in the chosen sample normally has three unblended detected lines with $E_{\rm up} > 400$\,K (419\,K, 526\,K and 782\,K) which in addition to the column density, allow the measurement of excitation temperature for the hot gas separately from that of the warm gas. However, when the 782\,K line is not detected the temperature is fixed to 300\,K. This value is chosen based on the mean of the hot CH$_3$CN excitation temperatures for other sources where a measurement is possible (see Appendix \ref{sec:add_tab_plots}). To test whether this assumption is valid, we varied the excitation temperature from 200\,K to as high as 500\,K for a few sources and found that the column densities could only change by a factor of $\lesssim2$. Therefore, the conclusions of this work will not be affected by this assumption. For the most line-rich sources the line with $E_{\rm up} = 931$\,K is also detected. When fitting these high-$E_{\rm up}$ lines, the lines with $E_{\rm up} < 400$\,K are ignored as they are usually (marginally) optically thick and even if they were not optically thick, their emission would be dominated by the warm component. Appendix \ref{sec:add_tab_plots} presents an example of the fitting for the hot column density of CH$_3$CN.

Methanol (CH$_3$OH), similar to methyl cyanide, normally has multiple relatively unblended lines with $E_{\rm up} > 400$\, detected in the chosen ALMAGAL sample. These lines have $E_{\rm up}$ of 508\,K, 746\,K, 776\,K, 802\,K, and 1181\,K.  The line with $E_{\rm up} = 508$\,K for the most line-rich sources becomes optically thick. Therefore, we ignore this line for all sources to avoid any potentially optically thick lines. When the line with ${\sim}1200$\,K ($\nu = 220.887$\,GHz and $A_{\rm ij} = 7.5 \times 10^{-5}$\,s$^{-1}$) is detected the excitation temperature is measured. Otherwise, it is fixed to 300\,K or the upper limit on temperature. Again the lines with $E_{\rm up} < 400$\,K are ignored in the fitting for robust measurements of the hot column densities.

Isocyanic acid (HNCO) in the chosen sample normally has two detected lines with $E_{\rm up} > 400$\,K, at $E_{\rm up} = 448$\,K and $E_{\rm up} = 750$\,K. We find the excitation temperatures in addition to the column densities for those sources. However, we note that these excitation temperatures are uncertain given the lack of detected lines with a large range of $E_{\rm up}$, while column densities can be measured accurately. If the HNCO line with $E_{\rm up} = 750$\,K is not detected toward a source the temperature is fixed to 300\,K. There are also two HNCO transitions with $E_{\rm up}$ of 1050\,K and 1450\,K which are useful for constraining the upper limit on excitation temperature. Given the tentative nature of C$_2$H$_3$CN results, a description of our methods to obtain C$_2$H$_3$CN column densities is given in Appendix \ref{app:C2H3CN}.

\begin{figure*}[ht!]
\plotone{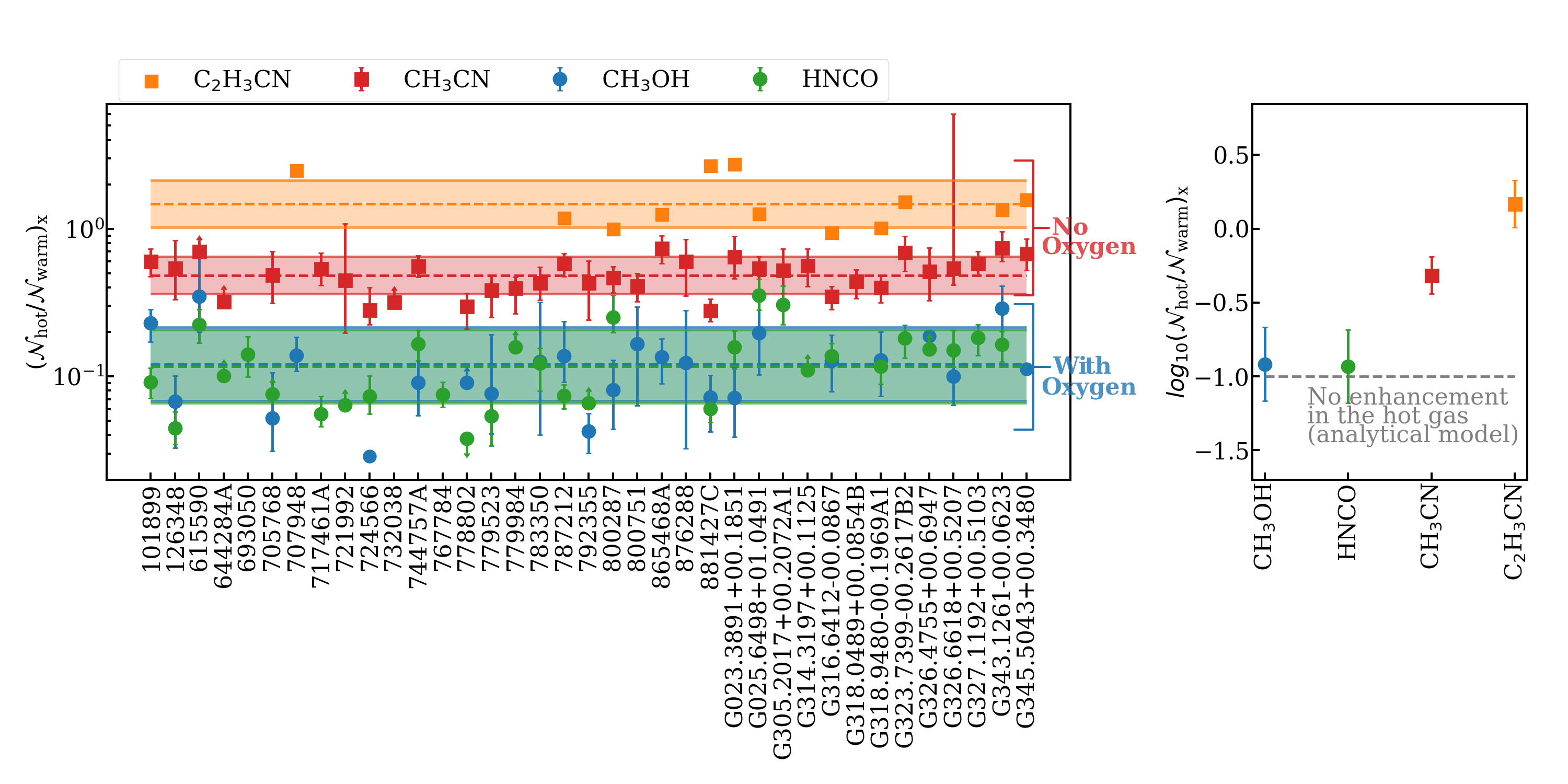}
\caption{Left panel is the same as Figure \ref{fig:ratio_methanol_methyl} but with the addition of C$_2$H$_3$CN (orange squares), and HNCO (green circles). For C$_2$H$_3$CN, given the data points are only approximate with no error bars, the non-weighted mean and standard deviation are shown. For all molecules the upper limits are eliminated before calculating the (weighted) mean and standard deviation. The warm components of HNCO and C$_2$H$_3$CN are taken from \cite{Nazari2022ALMAGAL}. The right panel presents the weighted (by the errors) mean of $log_{10}$ of the ratios of hot to warm number of molecules with the weighted standard deviation of $log_{10}$ of the data points for each molecule as the error bars. The hot to warm ratios of molecules without oxygen are at least ${\sim} 0.7$ dex higher than those with oxygen.
\label{fig:ratio_all}}
\end{figure*}

\section{Observational comparison of hot and warm components}
\label{sec:obs_comparison}

The total number of hot to the total number of warm molecules of methanol and methyl cyanide are presented in Figure \ref{fig:ratio_methanol_methyl}. The hot to warm ratios of methanol have a weighted mean of 0.12 while that for methyl cyanide is 0.48. The spread around the mean is tight for the two molecules with only a spread of a factor 1.78 and 1.33 around the mean for methanol and methyl cyanide, respectively. Figure \ref{fig:ratio_methanol_methyl} points to an enhancement in the hot component of an oxygen-poor molecule relative to an oxygen-rich one.

To examine whether this trend holds for other molecules with and without oxygen we also found the total number of hot molecules of HNCO (another molecule in addition to methanol with oxygen) and C$_2$H$_3$CN (another molecule in addition to methyl cyanide without oxygen) where possible for our sources (see Table \ref{tab:results}). Left panel of Figure \ref{fig:ratio_all} presents the hot to warm ratios of these molecules in addition to methanol and methyl cyanide. Figure \ref{fig:ratio_all} shows that HNCO has a similar mean and a similar standard deviation to methanol (factor of 1.77 around the mean). Moreover, the number of hot to the number of warm vinyl cyanide molecules are more similar to those of methyl cyanide with a mean of 1.47 and a factor of 1.44 scatter around the mean. This results in two distinct population of molecules, with at least a ${\sim}0.7$ dex (factor of ${\sim}$5) difference between the mean of the ratios. To conclude, we find that oxygen-poor molecules have a stronger contribution from the hot gas relative to oxygen-rich ones.

\section{Envelope-only analytical model}
\label{sec:toy}

Here we explain what is expected for the hot and warm number of molecules from a simple analytical model. Detailed radiative transfer models of low- and high-mass protostars and the effect of temperature structure on number of molecules are found in \cite{Nazari2023}.

The number of molecules of species $i$ in the gas phase is the abundance of that molecule ($X_{i}$) multiplied by the total number of hydrogen atoms in the gas phase. Here we assume that $X_{i}$ is solely found from the sublimation of ices at temperature $T_{\rm sub}$. Hence, the total number of molecules for species $i$ in the gas phase is

\begin{equation}
     \mathcal{N}_{i} = \int_{0}^{R_{T_{i, \rm sub}}} 4 \pi X_{i} n_{\rm H}(R) R^{2}dR 
    \label{eq:COM_num}
\end{equation}

\noindent where $R$ is the radius, $n_{\rm H}$ is the number density of hydrogen nucleus and $R_{T_{i,\rm{sub}}}$ is the radius at which the molecule $i$ is sublimated. Once a molecule is thermally sublimated into the gas phase it can trace the hot regions close to the protostar or warm regions further from the protostar at $R < R_{T_{\rm sub}}$.

Therefore, to measure the number of molecules in the hot and warm gas, Equation \eqref{eq:COM_num} can be re-written with the integral limit to be either $R_{T_{i,\rm{hot}}}$ or $R_{T_{i,\rm{warm}}}$ instead of $R_{T_{i,\rm{sub}}}$. Assuming a spherical model with a power-law density and temperature as a function of radius with powers $p=-1.5$ and $q=-0.4$ based on observations and radiative transfer models (\citealt{Adams1985}; \citealt{vanderTak2000}; \citealt{Nazari2022Lowmass}), the number of molecules for species $i$ in the hot (or warm) gas can be written as (see Appendix B of \citealt{Nazari2021})

\begin{equation}
    \mathcal{N}_{i, \rm hot} \propto X_{i, \rm hot} T_{i,\rm{hot}}^{-3.75},
    \label{eq:N_i}
\end{equation}

\noindent The total number of molecules of species $i$ in the hot gas and in the warm gas can then be compared as

\begin{equation}
    \frac{\mathcal{N}_{i, \rm hot}}{\mathcal{N}_{i, \rm warm}} = \frac{X_{i,\rm hot}}{X_{i,\rm warm}} \left(\frac{T_{i,\rm{hot}}}{T_{i,\rm{warm}}}\right)^{-3.75}.
    \label{eq:ratio}
\end{equation}

Assuming $T_{\rm hot} = 300$\,K and $T_{\rm warm} = 150$\,K consistent with the average excitation temperatures of the hot and warm gas for the studied molecules (see Appendix \ref{sec:add_tab_plots}), $(T_{i,\rm{hot}}/T_{i,\rm{warm}})^{-3.75} =0.07$. Therefore, if there is no additional gas phase chemistry in the hot gas (i.e., $X_{i,\rm hot}/X_{i,\rm warm} = 1$), 
\begin{equation}
    \frac{\mathcal{N}_{i, \rm hot}}{\mathcal{N}_{i, \rm warm}} \sim 0.1.
\end{equation}

\noindent It is worth noting that this result is valid for molecules with sublimation temperatures of $\lesssim 100-150$\,K so that there exists a warm component in the gas. For example if a molecule has a sublimation temperature of $200-300$\,K (e.g., NH$_2$CHO) the molecule will not be present in the gas-phase at 150\,K and this result will not be valid for that molecule. The molecules considered in this work (CH$_3$OH, CH$_3$CN, HNCO and C$_2$H$_3$CN) have similar binding energies and hence similar sublimation temperatures at around 100-150\,K in the interstellar conditions (\citealt{Collings2004}; \citealt{Das2018}; \citealt{Song2016}; \citealt{Wakelam2017}; \citealt{Bertin2017}; \citealt{Penteado2017}; \citealt{Ferrero2020}; \citealt{Busch2022}; \citealt{Minissale2022}).

There are some caveats in the above calculation, such as the assumed power of -0.4 for the temperature structure as a function of radius. This value is based on the regions where the envelope is optically thin to the reprocessed emission. To check how this assumption changes the calculations we take a steeper temperature profile with $q=-0.5$ for the optically thick regime (see \citealt{Adams1985} for low-mass and \citealt{Nazari2023Highmass} for high-mass sources). This will change the power of -3.75 to -3.0 and the value of $\mathcal{N}_{i, \rm hot}/\mathcal{N}_{i, \rm warm}$ from 0.07 to 0.13. Therefore, our assumption of $q$ for high-mass sources such as those from the ALMAGAL sample is only expected to produce a small scatter in the value of 0.1. Moreover, the larger the difference between the assumed $T_{\rm hot}$ and $T_{\rm warm}$, the smaller would be the ratio of $\mathcal{N}_{i, \rm hot}/\mathcal{N}_{i, \rm warm}$. The conclusion is that without any gas-phase chemistry the number of molecules in the hot gas should be around one order of magnitude smaller than that in the warm gas with some variations (less than a factor 2) based on the exact thermal and density structure of the protostellar system.

\section{Discussion and conclusions}
\label{sec:discussion}

Our findings in Section \ref{sec:obs_comparison} show that the mean of hot number of molecules to warm number of molecules for oxygen-rich species agrees well with the expected ratio of ${\sim} 0.1$ from the simple analytical model in Section \ref{sec:toy}. However, this ratio for oxygen-poor species deviates from the analytical model by a factor of about 5 (right panel of Figure \ref{fig:ratio_all}). Therefore, the main assumption made in that model, that is no additional gas-phase chemistry in the hot gas, does not hold for oxygen-poor molecules. This means an enhancement of oxygen-poor species in the hot gas and $X_{i,\rm hot}/X_{i,\rm warm}$ of ${\gtrsim}5$ for these molecules (see Equation \ref{eq:ratio}). In other words, there should be a gas-phase formation route for methyl cyanide (or vinyl cyanide) in the hot gas that does not exist for methanol (or isocyanic acid). This additional route could be the top-down formation mechanism suggested by \cite{vantHoff2020} for nitrogen-bearing oxygen-poor COMs as a result of destruction of refractory organics. 

When refractory organics are destroyed nitrogen- and carbon-bearing compounds should increase in the hot gas more than the oxygen-bearing compounds. This is because of the higher percentage of nitrogen and carbon in cometary dust grains compared with cometary ices, while the percentage of oxygen in cometary dust grains is lower than in ices (see \citealt{Rubin2019} Table 5). Moreover, the fraction of nitrogen to carbon in pre-solar organic nano-globules is ${\sim}0.1$ and thus nitrogen exists in dust grains (\citealt{Jones2016}). The nitrogen- and carbon-bearing compounds could be in the form of NH$_3$, CN, CH$_4$ and other forms of hydrocarbons (\citealt{Nakano2003}; \citealt{wang2013effect}; \citealt{Gail2017}; \citealt{vantHoff2020}). This agrees well with chemical models of protoplanetary disks where they find an increase of nitrogen-bearing species such as HCN and depletion of oxygen-rich ones as a result of carbon grain destruction (\citealt{Wei2019}). This is also inline with the results from the Infrared Space Observatory which showed enhanced abundances of C$_2$H$_2$ and HCN with increasing temperature in the gas around massive young stellar objects (\citealt{Lahuis2000}). 

After this additional release of nitrogen- and carbon-bearing species due to the destruction of refractory organics, species without oxygen such as CH$_3$CN could form in the gas-phase. One possible formation route is reaction of CH$_3^+$ and HCN to first form CH$_3$CNH$^+$ which can then produce CH$_3$CN through electronic dissociative recombination (\citealt{Garrod2022}). At higher temperatures above 300\,K, atomic nitrogen is expected to be released from NH$_3$, NH$_2$ or NH that can then react with CH$_3$ to produce HCN which could again turn into CH$_3$CN through the same mechanism as explained above (\citealt{Garrod2022}). For C$_2$H$_3$CN, a plausible high-temperature gas-phase formation route could be the reaction of C$_2$H$_4$ with CN (\citealt{Garrod2022}). Both of these minor species are expected to be released from destruction of refractory organics.

We attribute the enhancement of oxygen-poor molecules, CH$_3$CN and C$_2$H$_3$CN, in the hot gas to destruction of refractory organics and find it to be common in high-mass protostellar systems. It is yet to be confirmed whether the trend seen in Fig. \ref{fig:ratio_all} holds for other oxygen-poor (e.g., C$_2$H$_5$CN) and oxygen-rich species (e.g., CH$_3$OCH$_3$). If this conclusion can be extended to low-mass protostellar systems (pending further observations), the carbon deficiency on Earth could be generalized to other terrestrial planets. 

The grains that are destroyed at high temperatures most likely have similar composition to cometary refractory material which have a higher percentage of carbon and nitrogen than oxygen (\citealt{Rubin2019}). Therefore, destruction of these grains in the hot gas will inevitably increase the C/O and N/O ratios in the hot gas close to the protostars compared to the warm gas further away. In other words, Figure \ref{fig:ratio_all} shows a systematic increase in the C/O and N/O ratios of the hot gas. However, the exact values of these ratios depend on the fraction of refractory organics that are destroyed in these high-temperature regions which is not known. Future chemical models to constrain these elemental ratios based on the abundances of minor carbon, nitrogen and oxygen carriers discussed in this work are needed to constrain the fraction of refractory organics that are destroyed in the hot gas. Assuming a few caveats the measured fraction from these chemical models can then be used to estimate the amount of carbon depletion in rocky planets forming in the inner regions of protostellar systems. These caveats include a similar destruction of grains in low-mass protostars, particularly in the inner parts of the disk, and a weak dependence of the inner disk gas and dust elemental abundances on dust traps.

%\begin{acknowledgments}
\section{Acknowledgments}
We thank the referee for the helpful and constructive comments. We thank A.~P. Jones and P. Caselli for the helpful discussions. Astrochemistry in Leiden is supported by EU A-ERC grant 101019751 MOLDISK, NOVA, and by the NWO grant 618.000.001. Support by the Danish National Research Foundation through the Center of Excellence “InterCat” (Grant agreement no.: DNRF150) is also acknowledged. B.T. acknowledges support from the Programme National ``Physique et Chimie du Milieu Interstellair'' (PCMI) of CNRS/INSU with INC/INP and co-funded by CNES. M.L.R.H. acknowledges support from the Michigan Society of Fellows. J.K.J. acknowledges support from the Independent Research Fund Denmark (grant number 0135-00123B). This paper makes use of the following ALMA data: ADS/JAO.ALMA\#2019.1.00195.L. ALMA is a partnership of ESO (representing its member states), NSF (USA) and NINS (Japan), together with NRC (Canada), MOST and ASIAA (Taiwan), and KASI (Republic of Korea), in cooperation with the Republic of Chile. The Joint ALMA Observatory is operated by ESO, AUI/NRAO and NAOJ. The National Radio Astronomy Observatory is a facility of the National Science Foundation operated under cooperative agreement by Associated Universities, Inc. 
%\end{acknowledgments}

%% To help institutions obtain information on the effectiveness of their 
%% telescopes the AAS Journals has created a group of keywords for telescope 
%% facilities.
%
%% Following the acknowledgments section, use the following syntax and the
%% \facility{} or \facilities{} macros to list the keywords of facilities used 
%% in the research for the paper.  Each keyword is check against the master 
%% list during copy editing.  Individual instruments can be provided in 
%% parentheses, after the keyword, but they are not verified.

% \vspace{5mm}
% \facilities{HST(STIS), Swift(XRT and UVOT), AAVSO, CTIO:1.3m,
% CTIO:1.5m,CXO}

%% Similar to \facility{}, there is the optional \software command to allow 
%% authors a place to specify which programs were used during the creation of 
%% the manuscript. Authors should list each code and include either a
%% citation or url to the code inside ()s when available.

% \software{astropy \citealt{2013A&A...558A..33A,2018AJ....156..123A},  
%           Cloudy \citealt{2013RMxAA..49..137F}, 
%           Source Extractor \citealt{1996A&AS..117..393B}
%           }

\appendix

\section{Spectroscopic information}
\label{sec:spec_info}

The line lists and spectroscopic information for CH$_3$CN are taken from the CDMS (\citealt{Kukolich1973}; \citealt{Boucher1977}; \citealt{Kukolich1982}; \citealt{Cazzoli2006}; \citealt{Muller2015}). The partition function has been calculated for temperatures below 500\,K in the CDMS and contribution from lower vibrational states have been included. The higher vibrational states each contribute to less than 1\% at 300\,K and hence are ignored here. If these were included the difference between methyl cyanide and methanol results presented here would be larger.

The spectroscopic data for CH$_3$OH are taken from the CDMS (\citealt{Lees1968}; \citealt{Pickett1981}; \citealt{Herbst1984}; \citealt{Xu1995}; \citealt{Xu2008}). The vibrational levels included in the partition function are sufficient for temperatures up to 300\,K. The line data for HNCO and C$_2$H$_3$CN are also taken from the CDMS (\citealt{Kukolich1971}; \citealt{Hocking1975}; \citealt{Stolze1985}; \citealt{Demaison1994}; \citealt{Colmont1997}; \citealt{Muller2008}). Low-lying vibrational modes are included in the calculation of partition function in the CDMS for C$_2$H$_3$CN.

\section{Measurement of hot column densities of C$_2$H$_3$CN}
\label{app:C2H3CN}

There are not enough lines with $E_{\rm up} > 400$\,K detected in the spectra, for robust measurement of hot column densities of C$_2$H$_3$CN. Usually there is only one relatively unblended line with $E_{\rm up} {\sim} 400$\,K in the spectra. Hence, we fix the temperature to 300\,K and only fit the line with $E_{\rm up} \sim 435.0$\,K to find an approximation for the hot column densities of C$_2$H$_3$CN. We do not measure the error bars as it is not possible to firmly measure them. These values are approximate and not as accurate as what we find for CH$_3$CN, CH$_3$OH and HNCO. Therefore, deeper observations are needed for robust measurement of the column densities of the hot component and those of the warm component from the $^{13}$C isotopologues of vinyl cyanide to avoid the potentially optically thick lines of C$_2$H$_3$CN with $E_{\rm up} < 400$\,K.

\section{Additional tables and plots}
\label{sec:add_tab_plots}

Table \ref{tab:results} presents the column densities and excitation temperatures of the hot component for the molecules and sources considered here. In this table a few column densities are not reported for various sources and molecules. For all our molecules if the upper limit on the excitation temperature of the hot gas is $\leq 200$\,K, the hot column density is not measured to avoid potential contribution from the warm gas in the calculations. 

Methanol in sources 721992 and 779984 does not have a column density because no line with $E_{\rm up}>600$\,K is detected and the minor isotopologue of methanol is also an upper limit in \cite{Nazari2022ALMAGAL}. For sources 707948, 865468A and G345.5043+00.3480 the high-$E_{\rm up}$ lines of HNCO are too blended and hence it is not possible to measure the hot column density. Lines of HN$^{13}$CO for source 800751 are too blended (\citealt{Nazari2022ALMAGAL}) and hence the column density for HNCO is not given here. For sources 732038 and 876288 only an upper limit for the hot component is possible while the $^{13}$C isotopologue is also an upper limit, therefore, those values are not useful for our analysis. For C$_2$H$_3$CN, the sources with no measurement are those with no detection of the warm component in \cite{Nazari2022ALMAGAL}.

\begin{figure*}[ht!]
\plotone{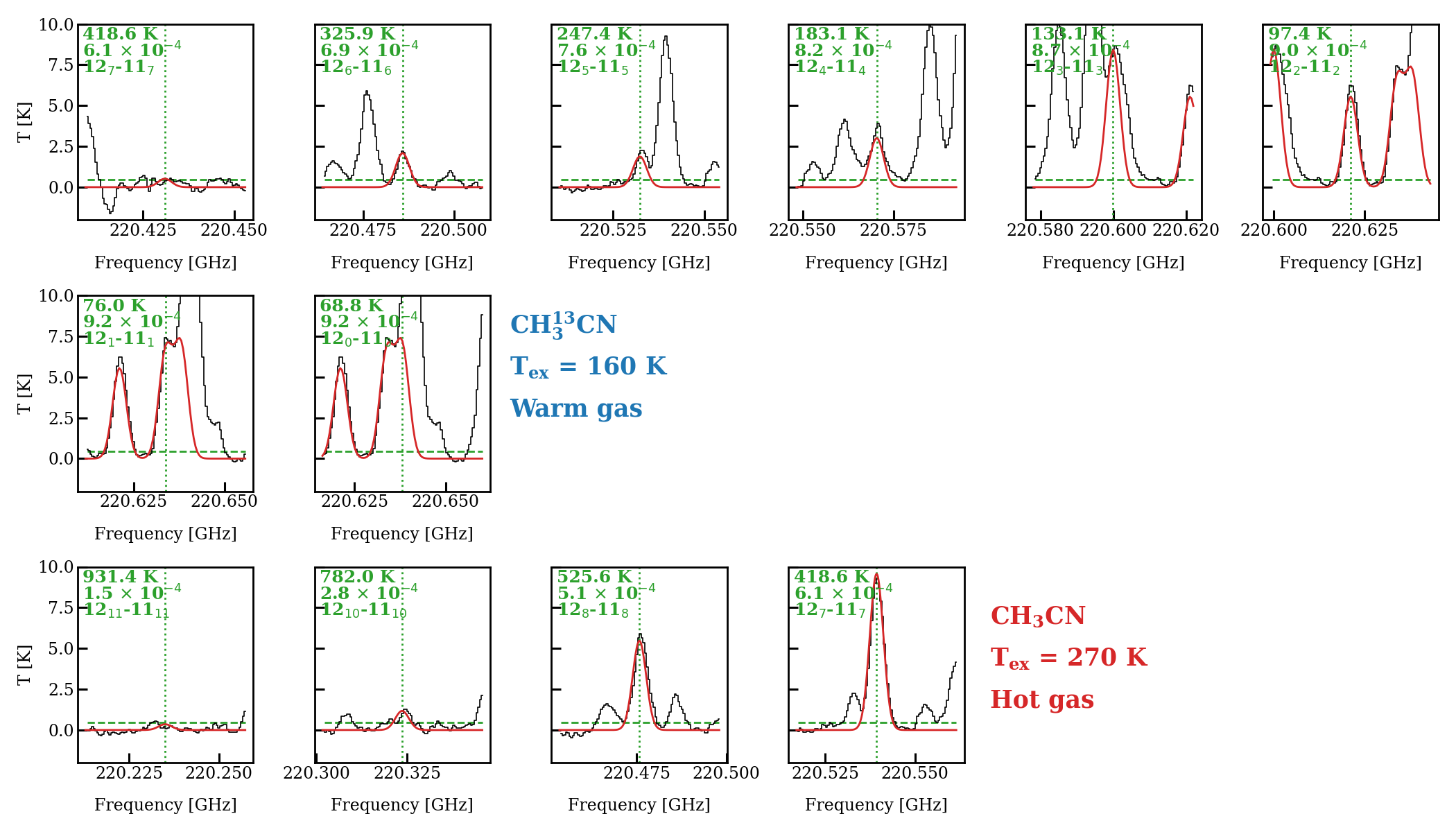}
\caption{Example of spectral fitting for the warm and hot gas for the relevant lines of CH$_3^{13}$CN and CH$_3$CN. Red shows the model and black shows the data. The top panels show the fit of CH$_3^{13}$CN for the source G316.6412-00.0867 (the best fit model is taken from \citealt{Nazari2022ALMAGAL}). The bottom panels show the fit of high-$E_{\rm up}$ lines of CH$_3$CN for the same source. The molecule names and measured excitation temperatures are printed on the right next to the panels. The upper energy levels ($E_{\rm up}$), $A_{\rm ij}$ coefficients and quantum numbers are printed in green on the top left of each panel. The green dashed lines show the $3\sigma$ level and the green vertical dotted lines highlight the transition frequency for each line. 
\label{fig:fitting}}
\end{figure*}

\begin{table*}
\Huge
\renewcommand{\arraystretch}{1.2}
    \caption{Column densities and excitation temperatures for the hot component.}
    \label{tab:results}
    \resizebox{\textwidth}{!}{\begin{tabular}{@{\extracolsep{2.5mm}}*{9}{l}}
          \toprule
          \toprule      
          & \multicolumn{2}{c}{CH$_3$CN}& \multicolumn{2}{c}{CH$_3$OH}&\multicolumn{2}{c}{HNCO} & \multicolumn{2}{c}{C$_2$H$_3$CN}\\
          \cmidrule{2-3} \cmidrule{4-5} \cmidrule{6-7} \cmidrule{8-9}
        Source & $N (\rm cm^{-2})$ &  $T_{\rm ex} (\rm K)$ &  $N (\rm cm^{-2})$ &  $T_{\rm ex} (\rm K)$ & $N (\rm cm^{-2})$ &  $T_{\rm ex} (\rm K)$ & $N (\rm cm^{-2})$ &  $T_{\rm ex} (\rm K)$ \\
        \midrule

101899 & 1.3$^{+0.1}_{-0.1}$ $\times 10^{16}$ & 230$^{+40}_{-40}$ & 3.0$^{+0.3}_{-0.3}$ $\times 10^{17}$ & 250$^{+20}_{-20}$ & 1.0$^{+0.1}_{-0.1}$ $\times 10^{16}$ & 320$^{+100}_{-90}$ & $<$\,2.5 $\times 10^{15}$ & [300] \\
126348 & 2.3$^{+1.1}_{-0.7}$ $\times 10^{15}$ & 350$^{+150}_{-170}$ & 7.5$^{+2.5}_{-2.0}$ $\times 10^{16}$ & 250$^{+50}_{-50}$ & 9.7$^{+2.3}_{-1.2}$ $\times 10^{14}$ & [300] & $<$\,1.2 $\times 10^{15}$ & [300] \\
615590 & 1.2$^{+0.1}_{-0.1}$ $\times 10^{16}$ & 250$^{+80}_{-80}$ & 6.5$^{+5.5}_{-2.3}$ $\times 10^{17}$ & 190$^{+30}_{-30}$ & 2.2$^{+0.3}_{-0.2}$ $\times 10^{16}$ & 310$^{+50}_{-50}$ & -- & -- \\
644284A & 3.7$^{+0.3}_{-0.2}$ $\times 10^{15}$ & [300] & -- &$<$\,190 & 6.5$^{+1.5}_{-0.8}$ $\times 10^{15}$ & 300$^{+140}_{-110}$ & -- & -- \\
693050 & -- & $<$\,160 & -- &$<$\,150 & 7.0$^{+2.0}_{-1.5}$ $\times 10^{15}$ & 210$^{+60}_{-50}$ & $<$\,8.5 $\times 10^{14}$ & [300] \\
705768 & 2.5$^{+1.0}_{-0.4}$ $\times 10^{15}$ & 350$^{+150}_{-190}$ & 9.0$^{+9.0}_{-3.0}$ $\times 10^{16}$ & 200$^{+50}_{-40}$ & 9.0$^{+2.0}_{-1.0}$ $\times 10^{14}$ & [300] & -- & -- \\
707948 & -- & $>$\,500 & 1.4$^{+0.4}_{-0.2}$ $\times 10^{18}$ & 250$^{+20}_{-30}$ & -- & -- & $\sim$ 2.0 $\times 10^{16}$ & [300] \\
717461A & 2.4$^{+0.2}_{-0.3}$ $\times 10^{15}$ & [300] & -- &$<$\,200 & 1.8$^{+0.2}_{-0.2}$ $\times 10^{15}$ & [300] & $<$\,9.0 $\times 10^{14}$ & [300] \\
721992 & 5.0$^{+7.0}_{-2.7}$ $\times 10^{15}$ & 120$^{+60}_{-30}$ & -- & -- & 6.0$^{+1.2}_{-1.0}$ $\times 10^{14}$ & [300] & $<$\,1.3 $\times 10^{15}$ & [300] \\
724566 & 8.0$^{+3.0}_{-0.8}$ $\times 10^{15}$ & 340$^{+110}_{-150}$ & $\sim$ 1.2 $\times 10^{17}$ & [300] & 3.0$^{+0.3}_{-0.2}$ $\times 10^{15}$ & [300] & $<$\,2.0 $\times 10^{15}$ & [300] \\
732038 & 3.0$^{+0.2}_{-0.3}$ $\times 10^{15}$ & [300] & -- &$<$\,190 & -- & -- & -- & -- \\
744757A & 4.0$^{+0.3}_{-0.2}$ $\times 10^{15}$ & 250$^{+80}_{-50}$ & 1.5$^{+1.3}_{-0.5}$ $\times 10^{17}$ & 210$^{+30}_{-40}$ & 4.8$^{+0.4}_{-0.4}$ $\times 10^{15}$ & [300] & $<$\,9.0 $\times 10^{14}$ & [300] \\
767784 & -- & $<$\,170 & -- &$<$\,180 & 1.6$^{+0.2}_{-0.1}$ $\times 10^{15}$ & [300] & -- & -- \\
778802 & 1.0$^{+0.1}_{-0.2}$ $\times 10^{15}$ & [300] & 3.0$^{+4.0}_{-1.0}$ $\times 10^{16}$ & 250$^{+90}_{-90}$ & $<$\,6.0 $\times 10^{14}$ & [300] & $<$\,1.0 $\times 10^{15}$ & [300] \\
779523 & 1.2$^{+0.1}_{-0.2}$ $\times 10^{15}$ & [300] & 1.0$^{+1.5}_{-0.4}$ $\times 10^{17}$ & 200$^{+50}_{-60}$ & 6.5$^{+1.0}_{-1.0}$ $\times 10^{14}$ & [300] & -- & -- \\
779984 & 1.0$^{+0.1}_{-0.2}$ $\times 10^{15}$ & [300] & -- & -- & 1.0$^{+0.1}_{-0.1}$ $\times 10^{15}$ & 430$^{+70}_{-80}$ & $<$\,7.5 $\times 10^{14}$ & [300] \\
783350 & 2.3$^{+0.2}_{-0.2}$ $\times 10^{15}$ & [300] & 8.0$^{+11.0}_{-3.2}$ $\times 10^{16}$ & 200$^{+60}_{-50}$ & 1.2$^{+0.1}_{-0.2}$ $\times 10^{15}$ & [300] & $<$\,8.0 $\times 10^{14}$ & [300] \\
787212 & 1.6$^{+0.1}_{-0.1}$ $\times 10^{16}$ & 230$^{+40}_{-30}$ & 4.5$^{+3.0}_{-1.0}$ $\times 10^{17}$ & 190$^{+20}_{-30}$ & 9.9$^{+0.1}_{-0.4}$ $\times 10^{15}$ & 350$^{+60}_{-80}$ & $\sim$ 2.7 $\times 10^{15}$ & [300] \\
792355 & 1.2$^{+0.1}_{-0.1}$ $\times 10^{15}$ & [300] & 3.0$^{+0.3}_{-0.3}$ $\times 10^{16}$ & [250] & 7.5$^{+1.5}_{-1.5}$ $\times 10^{14}$ & [300] & -- & -- \\
800287 & 1.3$^{+0.1}_{-0.1}$ $\times 10^{16}$ & [300] & 2.1$^{+1.0}_{-0.7}$ $\times 10^{17}$ & 240$^{+50}_{-50}$ & 8.0$^{+1.0}_{-0.8}$ $\times 10^{15}$ & 340$^{+160}_{-120}$ & $\sim$ 2.9 $\times 10^{15}$ & [300] \\
800751 & 3.4$^{+0.4}_{-0.4}$ $\times 10^{15}$ & 230$^{+40}_{-40}$ & 2.0$^{+0.9}_{-0.8}$ $\times 10^{17}$ & 180$^{+30}_{-20}$ & -- & -- & $<$\,7.5 $\times 10^{14}$ & [300] \\
865468A & 8.2$^{+1.3}_{-1.2}$ $\times 10^{16}$ & 430$^{+50}_{-50}$ & 1.7$^{+0.4}_{-0.4}$ $\times 10^{18}$ & 250$^{+30}_{-30}$ & -- & -- & $\sim$ 7.0 $\times 10^{15}$ & [300] \\
876288 & 2.2$^{+0.2}_{-0.2}$ $\times 10^{15}$ & [300] & 1.0$^{+1.1}_{-0.3}$ $\times 10^{17}$ & 190$^{+30}_{-40}$ & -- & -- & -- & -- \\
881427C & 2.6$^{+0.3}_{-0.2}$ $\times 10^{16}$ & 300$^{+40}_{-40}$ & 5.5$^{+1.0}_{-1.0}$ $\times 10^{17}$ & 250$^{+40}_{-30}$ & 1.2$^{+0.1}_{-0.1}$ $\times 10^{16}$ & 450$^{+50}_{-50}$ & $\sim$ 6.0 $\times 10^{15}$ & [300] \\
G023.3891+00.1851 & 3.5$^{+0.6}_{-0.5}$ $\times 10^{15}$ & 450$^{+50}_{-50}$ & 8.0$^{+4.0}_{-2.5}$ $\times 10^{16}$ & 250$^{+50}_{-50}$ & 1.8$^{+0.2}_{-0.1}$ $\times 10^{15}$ & 450$^{+50}_{-50}$ & $\sim$ 5.5 $\times 10^{14}$ & [300] \\
G025.6498+01.0491 & 1.0$^{+0.1}_{-0.1}$ $\times 10^{16}$ & 300$^{+50}_{-60}$ & 6.5$^{+3.5}_{-2.5}$ $\times 10^{17}$ & 190$^{+30}_{-30}$ & 1.0$^{+0.1}_{-0.1}$ $\times 10^{16}$ & 350$^{+100}_{-110}$ & $\sim$ 2.4 $\times 10^{15}$ & [300] \\
G305.2017+00.2072A1 & 6.0$^{+1.8}_{-1.5}$ $\times 10^{15}$ & 190$^{+40}_{-40}$ & -- &$<$\,200 & 4.7$^{+1.3}_{-0.9}$ $\times 10^{15}$ & 210$^{+50}_{-50}$ & $<$\,1.2 $\times 10^{15}$ & [300] \\
G314.3197+00.1125 & 5.5$^{+1.3}_{-1.2}$ $\times 10^{15}$ & 430$^{+70}_{-90}$ & -- &$<$\,200 & 2.3$^{+0.3}_{-0.3}$ $\times 10^{15}$ & [300] & $<$\,1.7 $\times 10^{15}$ & [300] \\
G316.6412-00.0867 & 1.1$^{+0.1}_{-0.1}$ $\times 10^{16}$ & 270$^{+50}_{-40}$ & 3.0$^{+1.3}_{-0.9}$ $\times 10^{17}$ & 230$^{+40}_{-30}$ & 6.0$^{+0.5}_{-0.5}$ $\times 10^{15}$ & 400$^{+100}_{-100}$ & $\sim$ 1.2 $\times 10^{15}$ & [300] \\
G318.0489+00.0854B & 3.9$^{+0.3}_{-0.4}$ $\times 10^{15}$ & [250] & -- &$<$\,170 & -- &$<$\,200 & $<$\,1.1 $\times 10^{15}$ & [300] \\
G318.9480-00.1969A1 & 1.8$^{+0.2}_{-0.1}$ $\times 10^{16}$ & 240$^{+40}_{-40}$ & 8.0$^{+4.0}_{-2.5}$ $\times 10^{17}$ & 210$^{+30}_{-30}$ & 1.7$^{+0.1}_{-0.2}$ $\times 10^{16}$ & 320$^{+80}_{-70}$ & $\sim$ 1.2 $\times 10^{15}$ & [300] \\
G323.7399-00.2617B2 & 1.2$^{+0.3}_{-0.2}$ $\times 10^{16}$ & 180$^{+40}_{-30}$ & -- &$<$\,190 & 1.0$^{+0.1}_{-0.2}$ $\times 10^{16}$ & 230$^{+30}_{-40}$ & $\sim$ 9.5 $\times 10^{14}$ & [300] \\
G326.4755+00.6947 & 1.1$^{+0.1}_{-0.1}$ $\times 10^{15}$ & [300] & $\sim$ 6.0 $\times 10^{16}$ & [220] & 1.2$^{+0.2}_{-0.2}$ $\times 10^{15}$ & [300] & $<$\,8.0 $\times 10^{14}$ & [300] \\
G326.6618+00.5207 & 2.8$^{+28.2}_{-0.4}$ $\times 10^{15}$ & 230$^{+60}_{-50}$ & 1.0$^{+0.6}_{-0.3}$ $\times 10^{17}$ & 200$^{+30}_{-30}$ & 3.1$^{+0.3}_{-0.3}$ $\times 10^{15}$ & [300] & $<$\,6.5 $\times 10^{14}$ & [300] \\
G327.1192+00.5103 & 1.5$^{+0.2}_{-0.1}$ $\times 10^{16}$ & [220] & -- &$<$\,200 & 2.1$^{+0.1}_{-0.1}$ $\times 10^{16}$ & 350$^{+100}_{-110}$ & $<$\,2.9 $\times 10^{15}$ & [300] \\
G343.1261-00.0623 & 2.2$^{+0.3}_{-0.2}$ $\times 10^{16}$ & 310$^{+50}_{-40}$ & 1.8$^{+0.7}_{-0.7}$ $\times 10^{17}$ & 220$^{+30}_{-30}$ & 2.6$^{+0.2}_{-0.2}$ $\times 10^{16}$ & 330$^{+90}_{-90}$ & $\sim$ 5.0 $\times 10^{15}$ & [300] \\
G345.5043+00.3480 & 4.2$^{+0.8}_{-0.7}$ $\times 10^{16}$ & 400$^{+60}_{-70}$ & $\sim$ 7.5 $\times 10^{17}$ & [250] & -- & -- & $\sim$ 7.0 $\times 10^{15}$ & [300] \\

\bottomrule
        \end{tabular}}
        \tablecomments{The excitation temperatures that are either fixed to 300\,K or to the upper limit on temperature are shown with square brackets.}
\end{table*}

\begin{figure*}[ht!]
\plotone{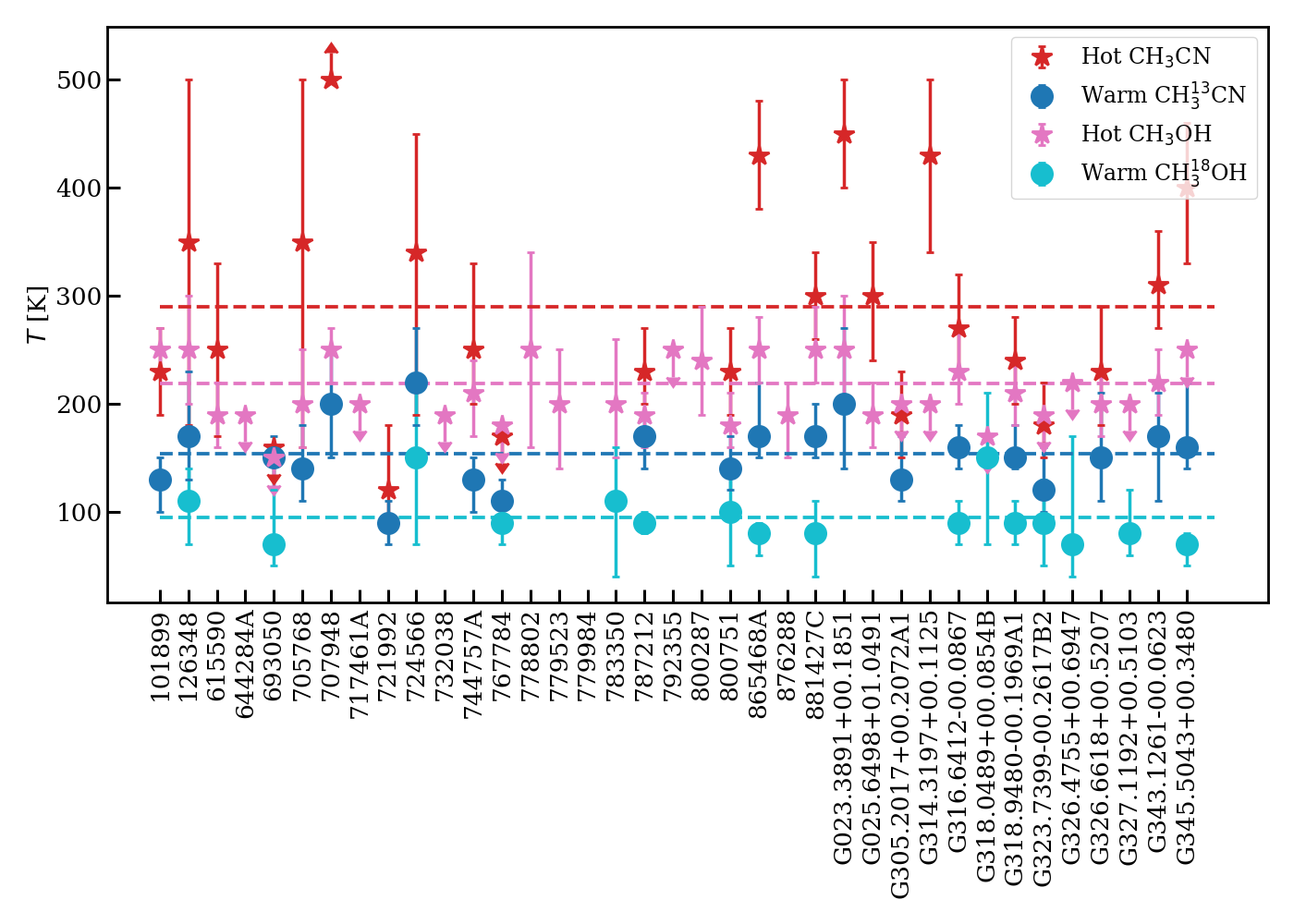}
\caption{The excitation temperatures of CH$_3$CN (red stars), CH$_3^{13}$CN (blue circles), CH$_3$OH (pink stars), and CH$_3^{18}$OH (cyan circles) where the measurement of excitation temperature was possible. The dashed lines show the mean for warm and hot methanol and methyl cyanide. The values for the $^{13}$C and $^{18}$O isotopologues of methyl cyanide and methanol are taken from \cite{Nazari2022ALMAGAL}. Given the large error bars the hot excitation temperatures of methanol and methyl cyanide are similar. The same can be said for their warm components.
\label{fig:T}}
\end{figure*}

Figure \ref{fig:fitting} presents an example source with the CASSIS fits (in red) on top of the data (in black) for measurement of the warm (top panels) and hot (bottom panels) column densities of methyl cyanide. Figure \ref{fig:T} presents the excitation temperatures of warm and hot methanol and methyl cyanide when the measurement was possible. Given the errors bars, the temperatures of hot methanol and methyl cyanide are similar, the same is true for the warm methanol and methyl cyanide.

%% For this sample we use BibTeX plus aasjournals.bst to generate the
%% the bibliography. The sample631.bib file was populated from ADS. To
%% get the citations to show in the compiled file do the following:
%%
%% pdflatex sample631.tex
%% bibtext sample631
%% pdflatex sample631.tex
%% pdflatex sample631.tex

\bibliography{CGD}{}
\bibliographystyle{aasjournal}

%% This command is needed to show the entire author+affiliation list when
%% the collaboration and author truncation commands are used.  It has to
%% go at the end of the manuscript.
%\allauthors

%% Include this line if you are using the \added, \replaced, \deleted
%% commands to see a summary list of all changes at the end of the article.
%\listofchanges

\end{document}